\documentclass[]{aastex631}

\usepackage{amssymb}
\usepackage{graphicx}
\usepackage{multirow}

\usepackage{footmisc}   %for multi-use of same footnote
\graphicspath{{Figures/}}
\usepackage{xcolor}
\usepackage{enumitem}

\usepackage{gensymb}
\usepackage{textcomp}

\hypersetup{pdfnewwindow=true, colorlinks=true,  linkcolor=magenta, citecolor=blue, filecolor=cyan, urlcolor=purple, final=true}
\received{June 7, 2024}
\revised{September 25, 2024}
\accepted{September 27, 2024}

\begin{document}
\title[Nested Rings CME Structure]{A Study on the Nested Rings CME Structure Observed by the WISPR Imager Onboard Parker Solar Probe} 

\correspondingauthor{Shaheda Begum Shaik}
\email{sshaik7@gmu.edu}

\author[0000-0002-3089-3431]{Shaheda Begum Shaik}
\affiliation{George Mason University, Fairfax, VA 22030, USA}
\affiliation{U.S.\ Naval Research Laboratory, Washington, DC 20375, USA}

\author[0000-0002-4459-7510]{Mark G. Linton}
\affiliation{U.S.\ Naval Research Laboratory, Washington, DC 20375, USA}

\author[0000-0001-9831-2640]{Sarah E. Gibson}
\affiliation{National Center for Atmospheric Research, Boulder, CO 80305, USA}

\author[0000-0003-1377-6353]{Phillip Hess}
\affiliation{U.S.\ Naval Research Laboratory, Washington, DC 20375, USA}

\author[0000-0002-3253-4205]{Robin C. Colaninno}
\affiliation{U.S.\ Naval Research Laboratory, Washington, DC 20375, USA}

\author[0000-0001-8480-947X]{Guillermo Stenborg}
\affiliation{The Johns Hopkins University Applied Physics Laboratory, Laurel, MD 20723, USA}

\author[0000-0003-1485-9564]{Carlos R. Braga}
\affiliation{The Johns Hopkins University Applied Physics Laboratory, Laurel, MD 20723, USA}

\author[0000-0001-6590-3479]{Erika Palmerio}
\affiliation{Predictive Science Inc., San Diego, CA 92121, USA}
%---------------------------------------------
\begin{abstract}

Despite the significance of coronal mass ejections (CMEs) in space weather, a comprehensive understanding of their interior morphology remains a scientific challenge, particularly with the advent of many state-of-the-art solar missions such as Parker Solar Probe (Parker) and Solar Orbiter (SO). In this study, we present an analysis of a complex CME as observed by the Wide-Field Imager for Solar PRobe (WISPR) heliospheric imager during  Parker's seventh solar encounter. The CME morphology does not fully conform with the general three-part density structure, exhibiting a front and core not significantly bright, with a highly structured overall configuration. In particular, its morphology reveals non-concentric nested rings, which we argue are a signature of the embedded helical magnetic flux rope (MFR) of the CME. For that, we analyze the morphological and kinematical properties of the nested density structures and demonstrate that they outline the projection of the three-dimensional structure of the flux rope as it crosses the lines of sight of the WISPR imager, thereby revealing the magnetic field geometry. Comparison of observations from various viewpoints suggests that the CME substructures can be discerned owing to the ideal viewing perspective, close proximity, and spatial resolution of the observing instrument.  

\end{abstract}
\keywords{Sun: coronal mass ejections (CMEs) -- Sun: solar corona -- Sun: heliosphere -- Sun: solar magnetic fields}

%---------------------------------------------
\section{Introduction}\label{Intro}

Over several decades, observations of coronal mass ejections (CMEs) have been facilitated by instruments on board spacecraft (S/C) such as the Solar and Heliospheric Observatory \citep[SOHO;][]{Domingo1995} and the Solar-Terrestrial Relations Observatory \citep[STEREO;][]{Kaiser2008}. These missions have enabled extensive investigations into the initiation mechanisms, internal morphology, and propagation characteristics of CMEs. While significant progress has been made in understanding the overall morphology and evolution of CMEs through these observations, there remains a persistent effort to discern many finer internal structures and their correspondence with magnetic flux ropes (MFRs) and filament material\footnote{Dense, chromospheric temperature plasma suspended in the corona in an elongated structure; named as filament when viewed on the solar disk and as prominence when viewed against the solar limb.}, if present. 

In chromospheric and coronal extreme ultra-violet (EUV), soft X-rays, radio, and white-light observations of the Sun in a pre-eruptive scenario, many studies have shown the presence of elliptical and dark coronal cavities around and above prominence material when viewed at the solar limb along the prominence axis \citep[e.g.,][and references within]{Hudson_2000, Marque_2004, Rachmeler_2013, Gibson_2015}. Coronal cavities are density-deficient regions that are considered to be the cross-section of MFRs, where the magnetic field strength is generally stronger compared to the surrounding coronal plasma \citep{LowH_1995, Rachmeler_2013}. Within their circular shape, these cavities have exhibited bullseye/nested patterns \citep{Urszula_2016} with spinning motion \citep{Wang_2010}, line-of-sight (LOS) counter-streaming flows, as well as plane-of-sky (POS) flows \citep{Schmit_2009, Urszula_2013}. It has been argued that these cavities indicate the presence of the pre-existing coronal flux ropes around the prominences extending along the line of sight \citep{Pneuman_1983, GibsonL_1998, Low_2001, Linker_2003}. They exhibit depleted emission, which indicates lower plasma density and higher magnetic flux than the ambient environment \citep{FullerG_2009}. When these coronal cavities erupt, they often become evident in the three-part (front-cavity-core) structure of CMEs \cite[as discussed in][]{GibsonF_2006, Vourlidas_2013, THoward_2017, Sarkar_2019, Hess_2020}. 

Since the launch of Parker Solar Probe \citep[Parker;][]{Fox2016}, the Wide-Field Imager for Solar Probe \citep[WISPR;][]{Vourlidas2016} has observed numerous CMEs in the inner heliosphere from remarkably closer distances to the Sun than had been previously achieved. This close proximity to erupting CMEs early in their evolution has allowed for the study of the physical properties and detailed substructures of these events \citep[e.g.,][]{Hess_2020, Wood_2021, Wood_2023, Howard_2022, Niembro2023}. In some cases, CMEs observed by Parker/WISPR have shown curved striations at the back of the CME, as shown in \cite{Howard_2022}---specifically Figures 5 and 6 within that study. 

In this work, we focus on a particular CME event that exhibited multiple circular rings of density enhancements that appear to be nested within each other in Parker/WISPR observations and which we call a ``nested ring'' structure. \cite{Chen_1997} and \cite{dere_1999} have suggested that the circular intensity patterns found in the internal structure of CMEs are signatures of a helical MFR forming the underlying structure of the CME. These studies were mainly based on observations of the SOHO/LASCO coronagraphs, which allow the coverage of the CME propagation within $30~R_\sun$. Later on, many works on CMEs observed in the visible spectral range have shown that a huge number of them exhibit a flux rope structure \citep[e.g.,][]{Wood_1999, KrallC_2006, Vourlidas_2013, Howard_2022}. However, not many of those studies show that a nested structure within the CME exists, nor have they addressed an in-depth investigation except in a few straightforward cases \citep[e.g.,][as mentioned above]{dere_1999, Chen_1997}.

The morphology of the CME under study here exhibits distinct sub-structures with apparently no pronounced bright front and core, which may not agree with the classical three-part CME structure \citep[e.g.,][]{IllingH_1985, CremadesB2004, Vourlidas_2013}. The observations suggest a three-dimensional MFR morphology exhibiting a non-concentric, nested ring structure that confines the whole CME. This event has been studied by \cite{Braga_2022}, who focused on the CME deformation during its evolution and the consequence of this deformation on estimating the time of arrival at 1~au. Through our investigation, we address the following scientific questions: (1) Are the nested rings a single, coherent structure? (2) Why do we see these nested ring structures? Are these part of the flux rope morphology? and (3) If all CMEs contain these nested ring structures, why are they not frequently observed through the available instruments?

This manuscript is structured as follows: Section~\ref{Section_2} provides the details of the CME initiation and the internal configuration of the CME. In Section~\ref{Section_3.1}, we examine the coherence of these nested rings, which shows whether the rings are one unified structure moving together or not. In Section~\ref{Section_3.2}, we discuss a proposed explanation for the presence of the nested structure, considering the morphology of the flux rope and the viewing geometry. The relative presence of the nested ring structure in the observations from different instruments is discussed in Section~\ref{Section_3.3}. Finally, in Section~\ref{Section_4}, we discuss and summarise the possible magnetic and density structures that could account for the appearance of such a nested ring structure from distinct viewing perspectives.

%-----------------------------------------------------------------------------
\section{Observations}\label{Section_2}

The event on 20--21 January 2021 was simultaneously observed in white-light imagery by the Parker/WISPR instruments as well as by the coronagraph instruments onboard the STEREO-A (hereafter ST-A) and the SOHO S/C. The Large Angle Spectroscopic Coronagraph \citep[LASCO;][]{Brueckner_1995} on board SOHO consists of three white-light coronagraphs with nested fields of view (FOVs): C1, C2, and C3, the one with innermost FOV (C1) being no longer operational. The two operational LASCO instruments observe the corona from the Sun--Earth Lagrange L1 point with overlapping FOVs of $1.5$ to $6~R_\odot$ and $3.7$ to $30~R_\odot$, respectively. From the Sun-Earth Connection Coronal and Heliospheric Investigation \citep[SECCHI;][]{Howard_2008} suite onboard ST-A, the SECCHI/COR2 white-light coronagraph observes from about 0.96~au, with a FOV covering heliocentric distances between ${\sim}2.5$ and $15~R_\odot$; and the Extreme Ultraviolet Imager \citep[SECCHI/EUVI;][]{Wuelser_2004} observes the full disk in EUV wavelengths, capturing chromospheric and coronal emission out to $1.7~R_\odot$ at four bandpass channels with peak responses at 171, 195, 284, and 304~{\AA}.

Parker/WISPR is a white-light heliospheric imager located on the ram side of the S/C  (i.e., facing the direction of motion of the S/C). It consists of two detectors with a combined, fixed angular FOV of about $95^\circ$ in the radial direction ($\sim13.5^{\circ}$--$53.5^{\circ}$ for the inner detector, hereafter WISPR-I; and ${\sim}50^{\circ}$--$108^{\circ}$ for the outer detector, hereafter WISPR-O). Due to the varying heliocentric distance of Parker, the angular extent of the FOV covers distinct projected solar distances as Parker approaches toward/recedes from perihelion \citep[see, e.g.,][]{Stenborg_2022}. The filter bandpass wavelengths of WISPR-I and WISPR-O are 490--790 nm and 475--795 nm, respectively. As white-light coronal instruments, ST-A/COR2, SOHO/LASCO, and Parker/WISPR record photospheric light scattered by the electrons in the solar corona and heliosphere via Thomson scattering \citep[i.e., the K-corona;][]{Billings1966} and by the interplanetary dust particles orbiting around the Sun \citep[i.e., the F-corona;][]{Grotrian1934, Kimura_1998}, along with the star field, Milky Way galactic sources, and planetary objects.

\begin{figure*}
\begin{center}
\includegraphics[width=.95\linewidth]{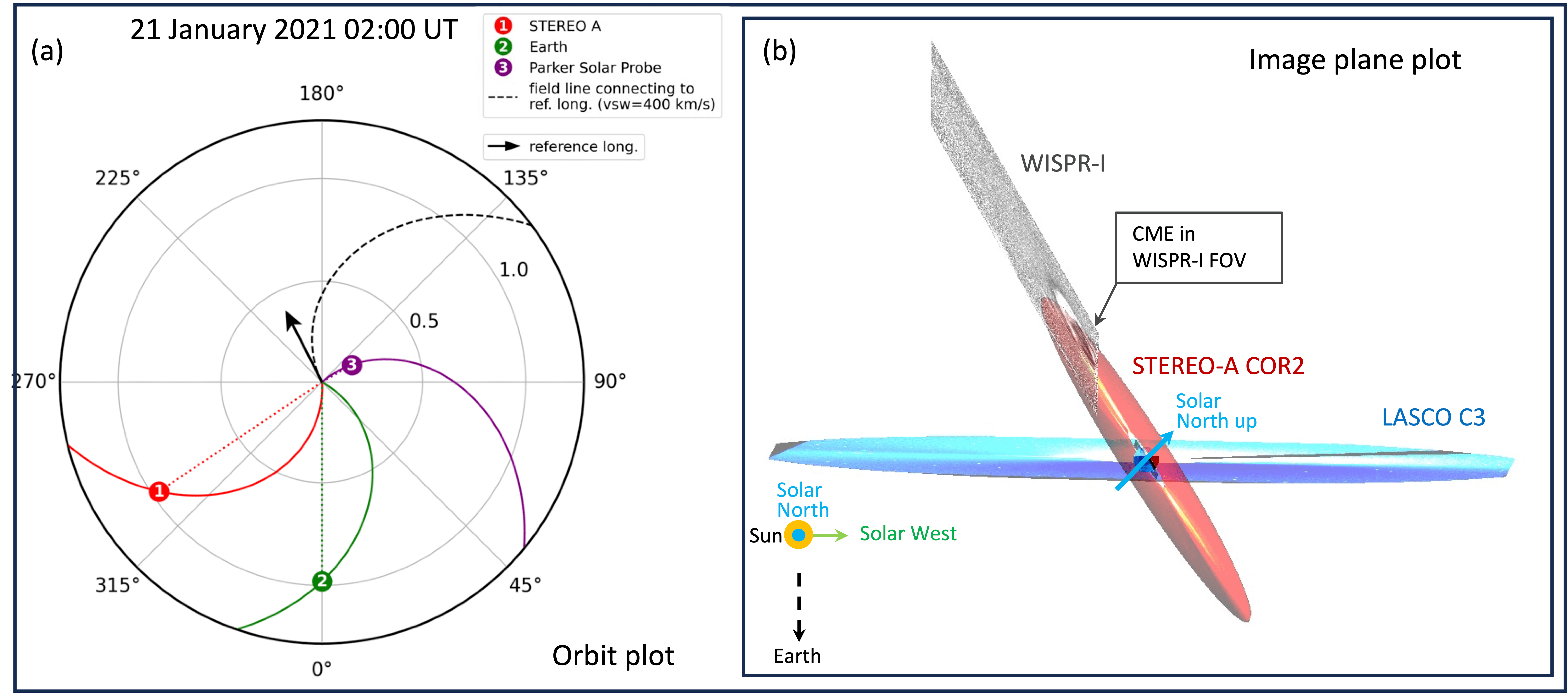}
\end{center}
\caption{Multi-spacecraft observations of the CME: (a) Orbit plot showing the respective locations of the S/C and Earth on 21 January 2021 at 02:00 UT (as numbered in the inset label). The black arrow points out the reference longitude of the CME, and the black dashed spiral shows a hypothetical heliospheric magnetic field line originating from that position. Radial distances at 0.5 and 1.0 au are delineated with the grey circles concentric on the sun. The angular information of the plot is given in Stonyhurst longitudes. (b) Image plane plot showing the corresponding image planes of each instrument (red: ST-A/COR2; grey: WISPR-I; blue: SOHO/LASCO-C3) as viewed from a perspective looking down on the ecliptic plane (similar to panel (a)), almost antiparallel to the direction of solar North up (marked by blue arrow nearly perpendicular to the page). The plot corresponds to a timestamp at around 02:12 UT on 21 January 2021, when the CME was in the FOVs of WISPR-I and ST-A/COR-2. The diagrams were produced using the Solar-MACH \citep{Gieseler2023} and JHelioviewer \citep{Muller2017} packages.} \label{Figure_1}
\end{figure*}

In Figure~\ref{Figure_1}(a), we display the locations of ST-A, Parker, and Earth on 21 January 2021 at 02:00~UT. The white-light observations consist of optically thin emission integrated along the LOS. Figure~\ref{Figure_1}(b) shows a snapshot of white-light images from each instrument (ST-A/COR2, LASCO/C3, and Parker/WISPR-I) as projected onto a plane that passes through the center of the Sun and perpendicular to the boresight of each camera, during the CME passage at around 02:12~UT on 21 January 2021. These image planes are displayed in the figure as viewed almost anti-parallel to the direction of the solar North pole up. Note that the image planes of WISPR-I and ST-A/COR-2 are parallel to each other as the CME crosses their FOVs. This is an optimal position to view the CME from opposite sides (i.e., separated by $\sim180^\circ$ in longitude), allowing for easy comparisons of the observed features.

%----------------
%---------------- 

\begin{figure}[ht]
\begin{center}
\includegraphics[width=.8\linewidth]{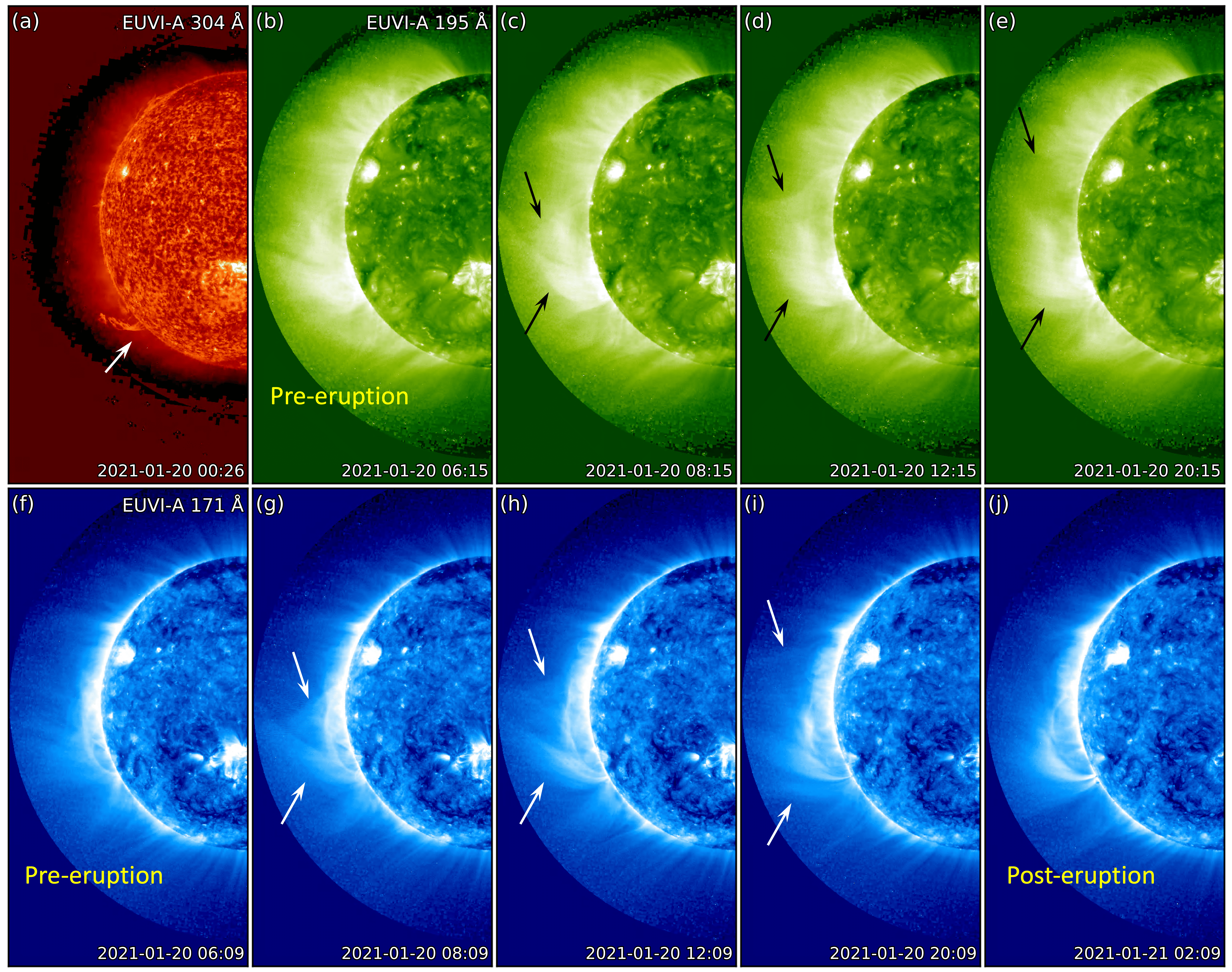}
\end{center}
\caption{Overview of the erupting activity off the solar eastern limb as observed by ST-A/EUVI in the (a) 304~{\AA}, (b--e) 195~{\AA} and (f--j) 171~{\AA} channels on 20 January 2021 (with times in UT). Panel (a), with a white arrow, shows a filament eruption that occurred a few hours before leading to the ``main'' eruption at around 08:00~UT. Panels (c--e) and (g--i) show the expansion and reconfiguration of magnetic field loops associated with the eruption off the eastern limb over the equator from the north and south hemispheres, whose outer edges are marked by arrows. Panels (b), (f), and (j) display the pre-eruptive and post-eruptive stages of the event, respectively.\\ (An animation of this figure is available.)} \label{Figure_2}
\end{figure}

Due to the CME--S/C geometry during the time of interest, the source region of the event remained on the far side of the Sun as observed from ST-A or SDO perspective (Figure~\ref{Figure_1}). Due to its apparent proximity to the limb, however, the reconfiguration of the overlying large magnetic field structure before and during the early stages of the eruption could be recorded in ST-A EUVI images.
In Figure~\ref{Figure_2}, we display snapshots of the pre-event scenario and the event as recorded in the 304 (panel a), 195 (panels b through e), and 171~{\AA} (panels f through j) channels of the ST-A/EUVI instrument. The full eruptive activity is shown in Supplementary Video~S1 of this manuscript. The eruption took place over an extended period of time (${\sim}$1 day) and as a two-step process. The event commenced gradually with the eruption of a filament in the southern hemisphere, as highlighted in Figure~\ref{Figure_2}(a). Panels (b) and (f) show the pre-eruptive stage of the event in the 195 and 171~{\AA} channels, respectively. Later on, after over seven hours, it developed as a large-scale streamer-blowout-like CME\footnote{A streamer blowout CME is an event that starts with the swelling of the overlying streamer followed by a CME, leaving the streamer significantly depleted.} \citep[e.g.,][]{Vourlidas_2018}, apparently originating from expanding large, bright magnetic loops on either side of the heliospheric current sheet as highlighted by black and white arrow marks in Figure~\ref{Figure_2}(c--e) and (g--i). The loops associated with the eruption had entirely faded by 21 January 2021, 02:09~UT, as shown in the post-eruptive stage in Figure~\ref{Figure_2}(j). 

%----------------
%----------------

%An animation shows the filament eruption in the ST-A/EUVI 304~{\AA} (left) channel, followed by eruptive activity and the associated magnetic field reconfiguration a few hours later, just behind the solar eastern limb, as observed in the 195 (middle) and 171~{\AA} (right) channels on 20--21 January 2021.

The CME was captured at coronal and inner heliospheric heights by the ST-A/COR1 and COR2, SOHO/C2 and C3, and Parker/WISPR instruments. The CME was visible in the ST-A/COR2 FOV on 20 January 2021 at $\sim$10:54~UT. Its development was observed until 21 January 2021 at  $\sim$06:00~UT, when the main structure of the CME completely moved out of the FOV. Although the event was observed from the opposite perspective compared to WISPR-I, the ST-A/COR2 observations do not show the detailed structure observed by WISPR-I (see Section~\ref{Section_3.3}). For the LASCO coronagraphs, it was a far-side event, as shown in Figure~\ref{Figure_1}. However, the front of the event was clearly seen developing in the combined FOV of LASCO/C2 from $\sim$14:30~UT on 20 January 2021 and C3 up to $\sim$06:00~UT on 21 January 2021. This allowed us to use the three sets of observations for the 3D reconstruction of the event (see Section~\ref{Section_3.2}).

Both the Parker/WISPR detectors observed this event with a cadence of $16$ minutes, starting on 20 January 2021 at $\sim$21:00 UT (WISPR-I; S/C at $\sim0.171$ au or $36.87~R_\sun$) and ending by 22 January 2021 at $\sim$15:00 UT (WISPR-O; S/C at $\sim0.227$ au or $49.01~R_\sun$). After crossing the outer boundary of WISPR-I, the CME emerged in WISPR-O, becoming fainter near the outer edge of the WISPR-O FOV. 

%------------------------------------------------------------------------
\subsection{CME Morphology and Development in the WISPR-I FOV}\label{sec:CMEmorph}

\begin{figure}[ht]
\begin{center}
\includegraphics[width=16.5cm]{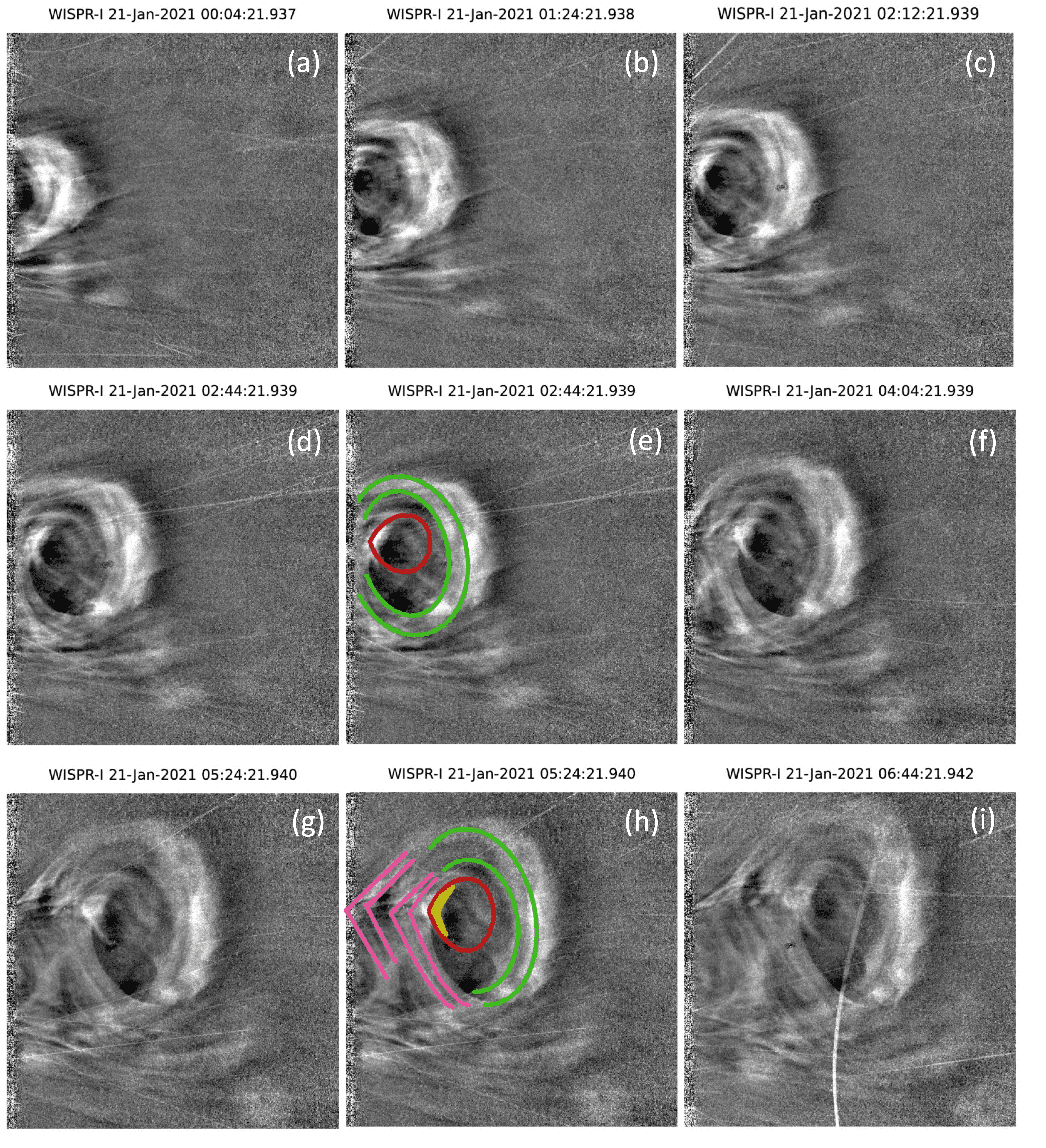}
\end{center}
\caption{A series of LW-processed WISPR-I detector images show the CME internal structure and its evolution on 21 January 2021 (with times in UT) with the Sun on the left of each image. The curves drawn manually in panels (e) and (h) represent the rings and other features that are observed as the CME propagates. The red curve indicates the inner core ring, the green curves are the two selected, clearly visible rings in the outer layer of the CME front, and the pink curves indicate the herringbone-shaped features at the backside of the core of the CME. The yellow curve within the core ring shows the bright inner base of the ring. Panels (d) and (e) display the WISPR image at the same time with and without the drawings, similarly for panels (g) and (h).}\label{Figure_3}
\end{figure}

Figure~\ref{Figure_3} displays the evolution of the CME during the first 6+ hours of its development in the WISPR-I FOV. After ${\sim}$06:00 UT on 21 January 2021 (not shown in the figure), the CME undergoes deformation and continues to deform as it propagates through the outer edge of the WISPR-I FOV and then through the WISPR-O FOV \citep[as discussed in][]{Braga_2022}. Note that panel pairs (d--e) and (g--h) display WISPR-I snapshots taken at the same time to allow the marking of certain CME features of interest by colored lines in panels (e) and (h). The images displayed have been processed with the algorithm described in Appendix~A of \cite{Howard_2022} (hereafter LW processing) to remove the F-corona signal and the contribution of pseudo-static K-corona features such as streamers and hence increase the relative contrast of the CME internal structure.

%------------------
%-----------------

A close visual inspection of Figure~\ref{Figure_3} reveals the following morphological features:

\begin{enumerate}
    \item A brightness pattern that exhibits at least three clearly discernible alternating bright and dark features in the main body of the CME, one innermost core ring, and two to three rings in the outer layer of the shell/CME; taken all together, we call this a ``nested ring structure'' (marked with the red and green curves in panels (e) and (h)).
    \item The shape of the core ring (pointed out by the red curve in panels (e) and (h)) appears to be independent of the outer rings. The inner base of this core ring is characterized by a bright `U' shaped feature (marked yellow in panel (h)).
    \item A herringbone pattern appears behind the core of the CME where the rings meet (marked pink in panel (h)). 
    \item Confined cavity regions are observed inside and outside the core ring. In the standard picture of the three-part CME structure, the inner cavity encompasses the axial field of the erupting magnetic flux rope with a direction normal to the page \citep{LowH_1995, Gibson_Low_2000}.
\end{enumerate}

Moreover, as the CME develops, the following changes are observed.
\begin{enumerate}
    \item All the rings seem to expand continuously as the CME propagates.
    \item The core ring seems to maintain its circular shape, although its front becomes irregular.
    \item Unlike the core ring, the overall shape of the outer rings becomes slightly oblong perpendicular to the direction of propagation (see panels (a) through (h)), whereas, in panel (i), they become flattened at the front and continue to deform thereafter. The thickness of the outer shell varies, and the rings contained within it seem to merge and separate at different points as they propagate.   
\end{enumerate}

%------------------
\begin{figure}[ht]
\begin{center}
\includegraphics[width=16.0cm]{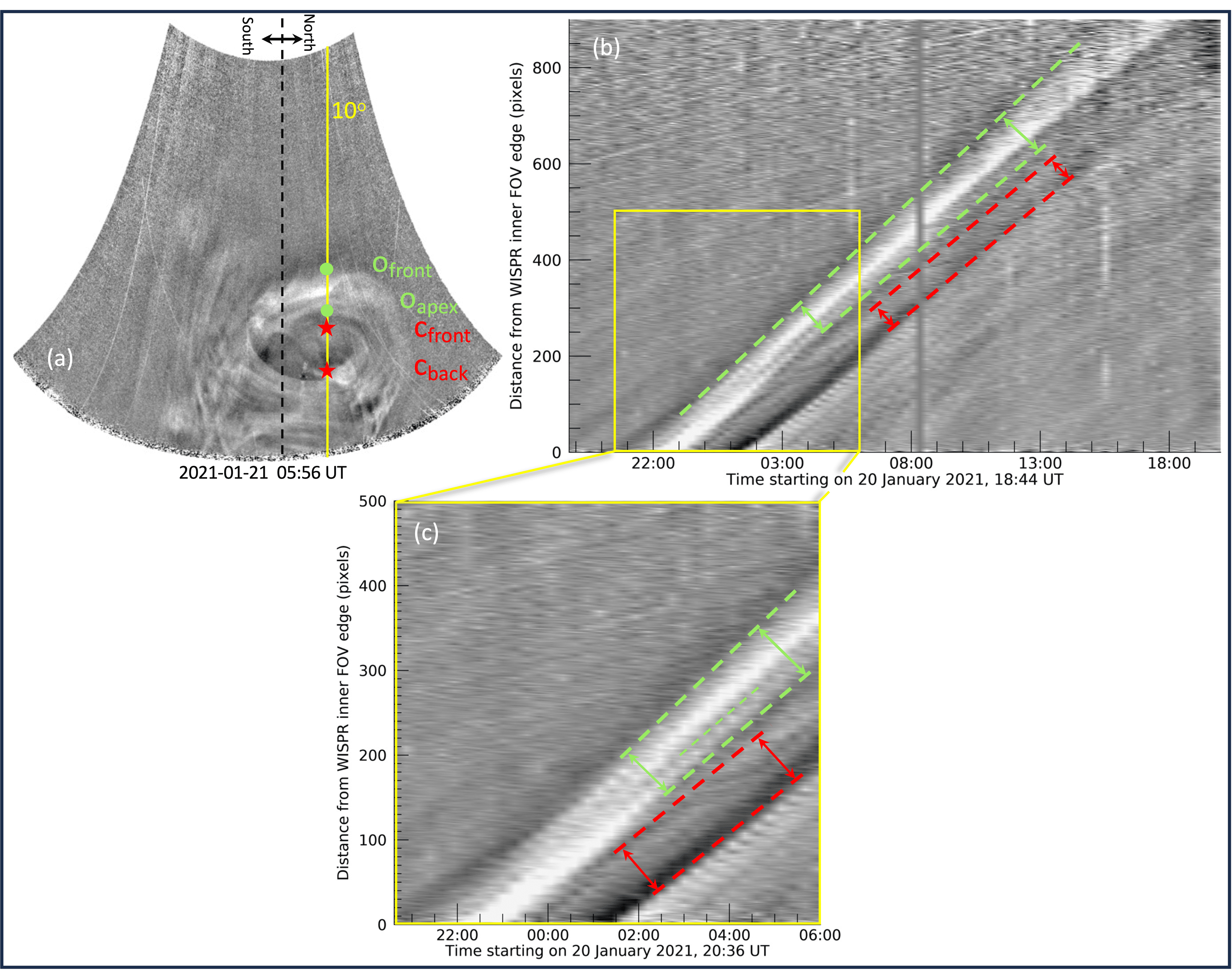}
\end{center}
\caption{Evolution of distinct features of the CME event. (a) Polar representation of a WISPR-I image at 05:56 UT on 21 January 2021, with the location of a radial slit at an angle of $10^\circ$ from the solar equator (black dashed vertical line) delineated with the yellow vertical line. The green and red symbols point out the features tracked on the outer and core rings, respectively. (b) Distance-time map of the development of the CME in the WISPR-I FOV. The evolution of the features tracked is marked with the same colors as in panel (a). The time period in which the features are clearly traceable is demarcated with the yellow box. (c) Zoomed in view of the distance-time map for the selected time period in the yellow box of panel (b). The thin dashed green line in panel (c) between $O_{front}$ and $O_{apex}$ marks another ring ($O^\prime_{front}$) within the outer shell, whereas the topmost dashed line of $O_{front}$ marks the second outer ring.}\label{Figure_4}
\end{figure}
%------------------
 
To evaluate this qualitative assessment, we first conduct a differential distance--time analysis. In particular, we placed linear slits along the selected radial rays from the Sun on the WISPR-I images to inspect the evolution of selected features along those directions. First, we transformed the Cartesian representation, [$x,y$], of the images into a polar representation, [$\theta, r$], where $\theta$ is the position angle (PA) of a particular radial direction (i.e., the angle relative to the projection of solar north, with vertex at the center of the Sun, measured counterclockwise in degrees) and $r$ is the distance from the center of the Sun (in the image plane, hence in pixels). For convenience, we rotate the reference system by $90^\circ$ clockwise so that $\theta$ is measured counterclockwise from the solar equator rather than from the solar north.
 
For illustration purposes, we show in Figure~\ref{Figure_4}(a) a polar projection of a WISPR-I image taken on 21 January 2021 at 05:56~UT. The vertical yellow line delineates the position of a radial slit at $\theta=10^\circ$ (the solar equator is at $\theta=0^\circ$, which is pointed out by the black dashed line). Along the radial slit, the green circles enclose the thickness of the outer rings between the front, $O_{front}$, and apex, $O_{apex}$ (backside of the ring at the leading edge as indicated in the Figure). The red symbols delimit the radial extent of the core ring, marked at the front, $C_{front}$, and the back, $C_{back}$ of the ring. Inspection of the CME evolution across the WISPR-I FOV in the polar representation shows that the flanks of the CME develop along $\theta=-22^\circ$ and $\theta=28^\circ$, corresponding to a constant angular extent of $\sim50^\circ$.

In Figure~\ref{Figure_4}(b), we display the distance--time map of the features that develop along the slit at $\theta=10^\circ$. Note that for representation purposes, we set the 0 pixels distance at the intersection of the selected radial slit with the inner edge of the FOV, hence the y-axis of the plot starts at 0 pixels. The dark and bright intensity features within the slit are seen as the inclined dark and bright tracks with positive slopes as the features propagate with time and outward from the Sun. The green and red dashed lines point out the tracks that describe the evolution of the four tracked features. The lengths of each dashed line marked are different to guide the reader. A zoomed-in view of the distance-time map in the yellow box of panel (b) is shown in panel (c) for the first few hours of time, during which the features are clearly traceable.

In panel (c), the slope of the outer ring at its front seems to differ from the slope at its apex by about $15\%$, which is indicated by different widths between the slopes, marked by green arrows at around 02:00 and 05:00~UT. As shown in panel (b), this trend remains for a longer time period, which suggests that the thickness of the outer shell increases over time (as indicated in our qualitative assessment listed above). After 01:00~UT, we start seeing in panel (c) a well-defined intensity track of another ring in the outer shell. To guide the eye, it is marked with a thin dashed line (starting at 03:00~UT). Later, after 06:00~UT, these tracks merge, showing one single strong-intensity track, which continues for a rather long time until fading away by around 13:00 UT in panel (b). The merging of the outer rings can be a consequence of the deformation that the CME undergoes after 06:00~UT.

The core ring appears to maintain its shape as it propagates. This is indicated by similar slopes of the core ring tracks shown by the red dashed lines at the front, $C_{front}$, and back, $C_{back}$ and, with equal widths between them at different times, as marked by the red arrows at around 02:00 UT and 05:00 UT. This also denotes that, as projected onto the image plane, the core ring continues to propagate with constant velocity at its front and back.
 
The structural changes revealed by the distance--time map could result from multiple factors, such as observation from a non-stationary S/C or the intrinsic characteristics of the CME and external effects from the ambient medium. If we assume an absolutely coherent CME expanding self-similarly as it develops, its structure would appear to change as it is being observed because of the rapid motion of the Parker S/C. In contrast, similar changes can also be observed if the internal structures of the CME move incoherently with varying velocities or if deformation occurs due to the inhomogeneous solar wind medium in which the CME propagates. To shed light on the coherence of this ring structure, we perform a more comprehensive analysis, including the characterization of the event kinematics and of its morphological evolution in 3D space.
 
%---------------------------------------------
\section{Nested Rings/MFR morphology}\label{Section_3}

\subsection{Coherence of the Nested Ring Structure}\label{Section_3.1}

CMEs, during their interplanetary propagation, may cease to be coherent magnetohydrodynamic structures \citep{Owens_2017} due to many factors, such as their interaction with the ambient solar wind and other CMEs, or streamers deflecting their trajectory. \cite{Braga_2022} found that the bulk of this CME develops self-similarly up to about 06:00~UT on 21 January 2021. Afterward, they argue that the CME starts to deform due to the interaction with the high-speed solar wind and other possible causes of loss of coherence and slow-mode shocks. Building upon their work, we investigate the coherence of the internal structure of the CME feature, examining whether the core and outer rings propagate as a single feature.

\begin{figure}[ht]
\begin{center}
\includegraphics[width=18cm]{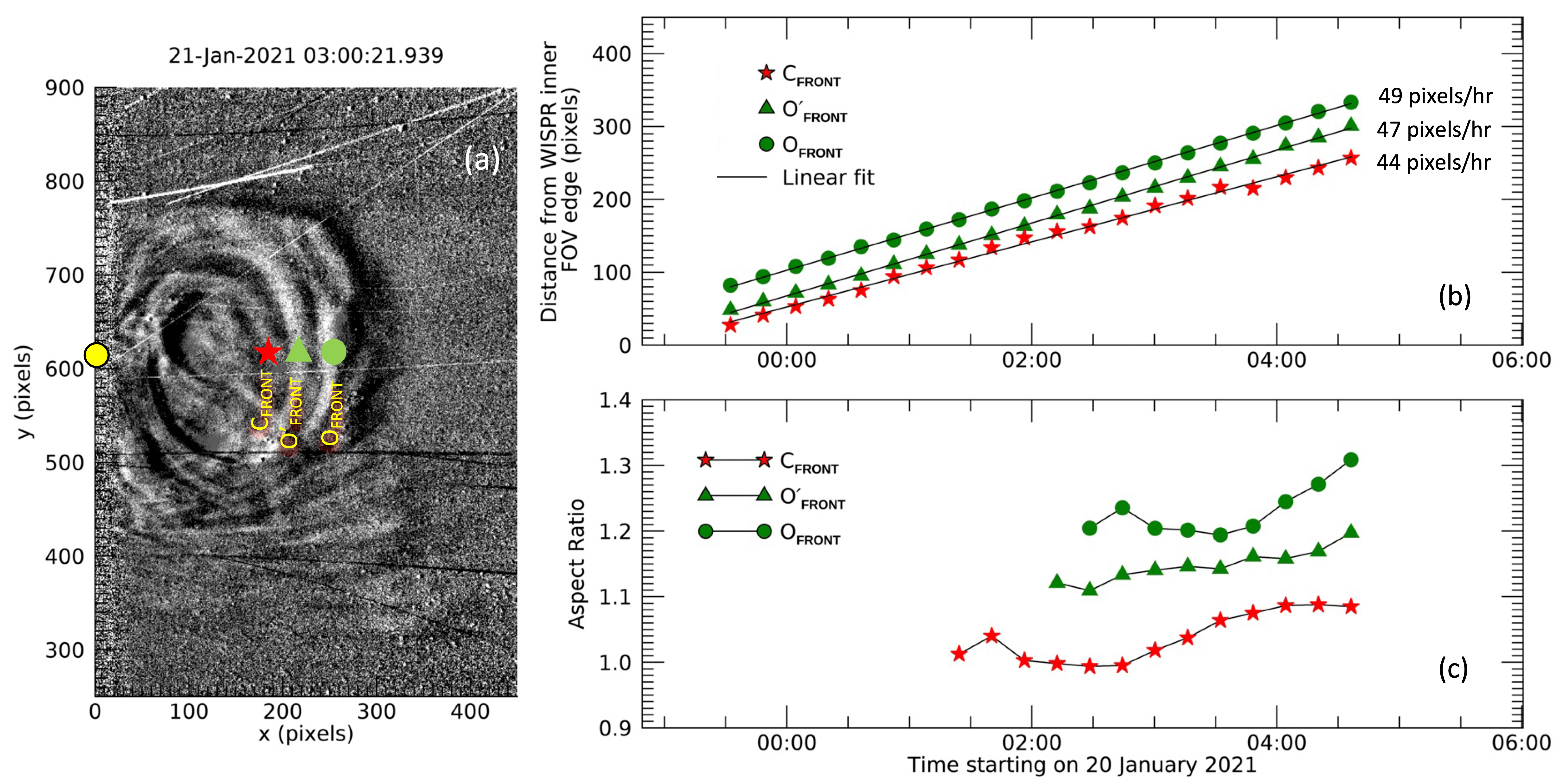}
\end{center}
\caption{Image plane analysis of the coherency of the nested rings: (a) Running difference image produced from the difference of the images at 03:00 UT and 02:44 UT on 21 January 2021. (b) Distance-time profiles and the evolution of the fronts of each ring ($C_{front}$, $O^\prime_{front}$, and $O_{front}$) are marked in panel (a) with star, triangle, and circle symbols. The linear fitting of the points is shown by the black line. (c) The aspect ratio (major-axis-length/minor-axis-length) of the rings from the times after which each of the rings becomes completely visible in the image FOV.}\label{Figure_5}
\end{figure}

The distance--time map analysis carried out in Section~\ref{sec:CMEmorph} showed the evolution of the selected features but without enough resolution for their kinematic analysis. Therefore, we conduct an analysis on the image plane using WISPR-I running-difference LW-processed images to further enhance the overall structure of the CME and associated faint features and track the fronts of the rings manually by pinpointing those locations (two outer rings' front, $O_{front}$, $O^\prime_{front}$, and core ring front, $C_{front}$). Figure~\ref{Figure_5}(a) shows a WISPR-I difference image between 21 January 2021 at 03:00 and 02:44~UT  with the colored symbols marking each of the tracked fronts ($O_{front}$ with a green circle, $O^\prime_{front}$ with a green triangle, and $C_{front}$ with a red star). A reference point, located where the outermost front first appears in the WISPR FOV at pixel coordinates [0,620], is marked with a yellow circle. In this analysis,  the distances reported for the respective fronts as they evolve are measured relative to this reference point (in pixels). The WISPR-I images after 04:36~UT are not considered for this analysis, as the specific points on the ring structure we aim to track become difficult to identify distinctly, as the deformation of the CME geometry continues to increase, especially after 06:00~UT.

%------------------
%-----------------

In Figure~\ref{Figure_5}(b), we display the distance--time plot for the three features tracked. The color code and symbols are as in panel (a). The linear fitting to the measurements of each feature is delineated with the continuous black lines. The two outer rings develop with a similar average projected speed in the image plane during the analyzed time period. The core ring, on the other hand, seems to have a similar development until 2:15~UT, after which the measurements seem to start vacillating around the linear trend suggested by the black line. Overall, and to a first approximation, the core ring seems to develop more slowly than the two outer rings, as indicated by the average speed values (in pixels/hr) depicted in the figure. This assessment is addressed in deeper detail in Section~\ref{sec:3D_trajectory}.

We also estimated the aspect ratio of the rings, which is computed by taking the ratio of the lengths of the major and minor axes of each of the rings. The evolution of the aspect ratio is displayed in Figure~\ref{Figure_5}(c), which shows that the shape of the rings becomes more elliptical as they develop. In addition, the center of the rings, as determined from the intersection point of the major and minor axes (not shown here), indicates that the rings appear to be non-concentric.

\subsubsection{Three-dimensional Trajectory and Coherence of the Ring Structure}\label{sec:3D_trajectory}

In the previous analysis, we manually determined the evolution of the ring structure as projected on the WISPR images, where each pixel records the integrated white-light emission along the LOS. This analysis does not provide information about the three-dimensional position and propagation of each of the rings observed. Therefore, we now extend the analysis by estimating the relative positions of the rings and their longitudes and latitudes with respect to the S/C, which allows us to estimate their physical speeds. CME geometry fitting techniques such as Fixed-$\phi$ \citep[F$\phi$;][]{Rouillard_2008} and Harmonic Mean \citep[HM;][]{Lugaz_2010} widely used with observations from the SECCHI Heliospheric Imagers \citep[HIs;][]{Eyles_2009} cannot be applied to Parker/WISPR images because they do not consider the change of the observer's location. On the other hand, \cite{Nindos_2021} characterize the development of the coronal transients in the WISPR FOV through the so-called J- and R-maps, assuming the features to be near the Thomson surface. \cite{Patel_2023}, in particular, compares the resulting speeds assuming the features with either an impact distance close to the Thomson surface (i.e., close to this surface) or close to the plane of sky.

Due to these caveats, to account for the changes in the longitude ($\phi_1$) and solar distance ($r_1$) of Parker during the evolution of the features of interest, we follow instead the methodology implemented in \cite{Braga_Vourlidas_2021}, which is based on the work of \citep[][and references therein]{Liewer_2020}, and is well-suited for a rapidly moving observer. In this approach, the assumption is made only on the radial movement of the features, and no assumptions are made on their longitude or latitude, which are self-consistently determined through the method, hence allowing to obtain their trajectory and speeds in 3D space. 

For each of the features tracked, at each time instance, we determine their angular elevation over the S/C orbital plane, $\beta_{obs}(t)$, and the elongation of its projection onto the orbital plane, $\gamma_{obs}(t)$ \citep[see, e.g., Figure 9 in][]{Braga_2022}. These angles are obtained by converting the Cartesian pixel locations of the features into the angular coordinates. For this purpose, we use the \texttt{wispr\_camera\_coords.pro} routine from the WISPR library in the IDL SolarSoft package \citep{Freeland_Handy_1998}.

%------------------
\begin{figure}[ht]
\begin{center}
\includegraphics[width=18cm]{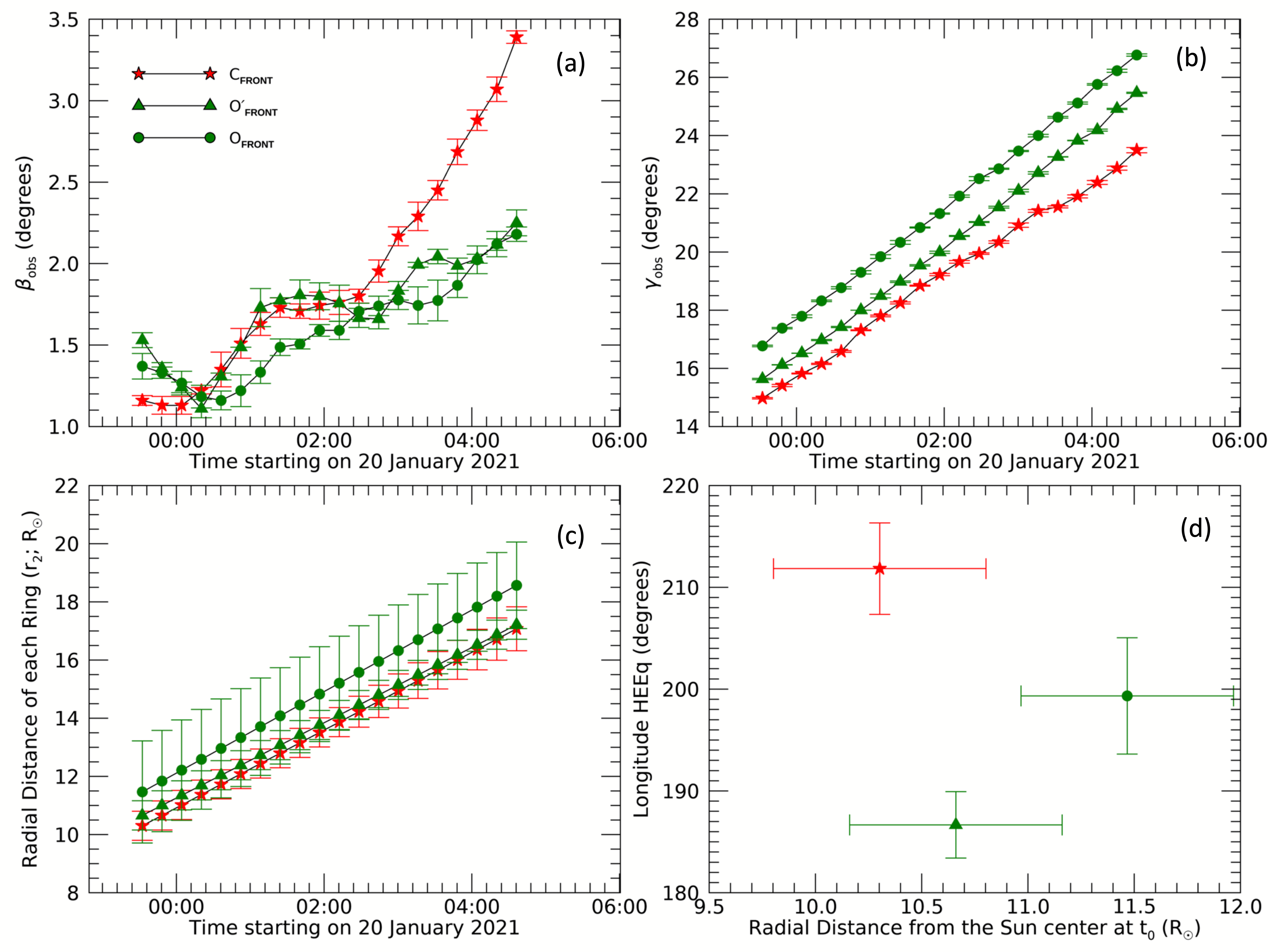}
\end{center}
\caption{Analysis of longitudes and radial distances for the coherency of the nested rings: (a) and (b) Evolution of image coordinates $\beta_{obs}$ and $\gamma_{obs}$ of the tracked features, i.e., the fronts of the three rings ($C_{front}$, $O^\prime_{front}$, and $O_{front}$). (c) Inferred radial distances in solar radii as a function of time for each of the rings. (d) Inferred longitudes of each of the ring fronts versus the radial distance plotted at the initial time of measurement, $t_0$. The symbols and color of the symbols and lines correspond to the three rings' fronts as used in Figure~\ref{Figure_5}(a).}\label{Figure_6}
\end{figure}
%-----------------

From the given set of $\beta_{obs}(t)$ and $\gamma_{obs}(t)$ angles with their vertex at the observer (i.e., Parker, which is located at Sun-centered spherical coordinates [$r_1$, $\phi_1$, $\delta_1$]), we determine the features' Sun-centered coordinates, [$r_2$, $\phi_2$, $\delta_2$], assuming that (1) their longitudes ($\phi_2$) and latitudes ($\delta_2$) do not change as they move radially outward from the solar center, and (2) the features develop with constant speed, $V$, in a heliocentric inertial (HCI) reference frame \citep{Burlaga_1984}, such that the features' radial coordinates evolve as $r_2(t) = r_2(t_0)+ V \cdot (t-t_0)$ with $t_0$ being the time of the initial measurement, which is taken as the reference time. As detailed in \cite{Liewer_2019}, to determine the unknowns, we solve the following equations for each feature:

\begin{equation} \label{eq1}
\frac{tan \beta (t)}{sin \gamma (t)} = \frac{tan\delta_2}{sin[\phi_2 - \phi_1(t)]},
\end{equation}
\begin{equation}\label{eq2}
cot \gamma(t) = \frac{r_1(t)-r_2(t)cos\delta_2\: cos[\phi_2-\phi_1(t)]}{r_2(t)cos\delta_2\: sin[\phi_2-\phi_1(t)]},
\end{equation}

With these two equations and a set of four unknown parameters $(r_2(t_0), \phi_2, \delta_2, V)$, we chose the set that results in a minimum difference between the time series of $\beta_{obs}(t)$ and $\gamma_{obs}(t)$ and $\beta(t)$ and $\gamma(t)$ given back from Equations~\ref{eq1} and \ref{eq2}. Additional details about the determination of these parameters are explained in \citet{Liewer_2019, Liewer_2020} and \citet{Braga_Vourlidas_2021}.

%---------------
\begin{table}[!ht]
\caption{Trajectory parameters of the leading edge of the rings}
%\centering
\begin{tabular}{c|c|c|c} 
\hline 
\rule{0pt}{3ex}

Trajectory Parameters& Core ring, $C_{front}$ & Outer ring 1, $O^\prime_{front}$ & Outer ring 2, $O_{front}$\\ 
\hline \hline
\rule{0pt}{3ex}

Longitude (HEEq), $\phi_2$ (degrees) & 211.8 $\pm$ 4.5 & 186.7 $\pm$ 3.3 & 199.3 $\pm$ 5.7\\ 
\rule{0pt}{3ex} 

Latitude (HEEq), $\delta_2$ (degrees) & 9.7 $\pm$ 2 & 6.4 $\pm$ 2 & 7.4 $\pm$ 2\\ 
\rule{0pt}{3ex}

Radial Distance, $r_2$ ($R_\odot$) at $t_0$& 10.3 $\pm$ 0.5 & 10.6 $\pm$ 0.5 & 11.5 $\pm$ 0.5 \\ 
\rule{0pt}{3ex}

\hspace{101pt} at $t_f$ & 17.0 $\pm$ 0.7 & 17.2 $\pm$ 0.5 & 18.5 $\pm$ 1.4 \\ \vspace{6pt}
\rule{0pt}{3ex}

Average Speed, V (${km\,s^{-1}}$)$^*$ & 250.1 $\pm$ 25  & 258.3 $\pm$ 25 & 270.8 $\pm$ 25\\
\hline
\end{tabular}
\label{table1}
\tablecomments{The iteration steps to solve equations~\ref{eq1} and \ref{eq2} are implemented every $2^\circ$ in longitude and latitude, 0.5 $R_\odot$ in radial distance, and 25 ${km\,s^{-1}}$ in speed. The uncertainty of the reported values was estimated as the standard deviation of the output parameters as obtained by repeating the analysis six times (i.e., by determining $\beta_{obs}(t)$ and $\gamma_{obs}(t)$ six different times). When the standard deviation is smaller than the step size of the iterative procedure, the step size is reported as the uncertainty. $t_f$ represents the time at which the final measurements were taken.} 
  %The parameters that have the obtained errors significantly smaller than the associated step sizes are listed with step sizes and with the error limits otherwise.}
\end{table}
%-----------------

Following this methodology, we track the fronts of the three rings ($C_{front}$, $O^\prime_{front}$, and $O_{front}$ as in Figure~\ref{Figure_5}). The obtained results are presented in Figure~\ref{Figure_6} and summarized in Table~\ref{table1}. To help track the CME features from Earth's perspective, the derived longitudes in the HCI system were converted to the Heliocentric Earth equatorial (HEEq) system \citep{Thompson_2006}, which is sometimes referred to as the Stonyhurst heliographic coordinate system. Since the time period under study is only five hours (from 20 January 2021, 23:30~UT to 21 January 2021, 04:36~UT), the change of the HEEq longitude relative to the fixed HCI longitude is $\sim 0.3^\circ$ and hence less than the errors range. Panels (a) and (b) show $\beta_{obs}(t)$ and $\gamma_{obs}(t)$ as measured for each feature (color and symbol code as indicated in the inset labels) for the analyzed time period. In panels (c) and (d), we show two prime output parameters of the analysis: the evolution of the radial distances and the longitude of each of the fronts of the rings, respectively. 

The time evolution of the radial distances of each of the rings in panel (c) exhibits a similar trend (similar slopes) in the time period analyzed. As reported in Table~\ref{table1}, the derived speeds seem to suggest only a slight variation between them. Their differences, however, are within the error range imposed by the constraints of the methodology. As per the assumptions of the methodology, the features are assumed to develop strictly radially, i.e., at a constant longitude and latitude along their evolution. Therefore, in Figure~\ref{Figure_6}(d), we display the derived longitudes for the three rings as a function of their respective radial distances at the time of their initial measurement, $t_0$. In the plot, and as also seen in Table~\ref{table1}, we note that there also exist slight variations in the derived longitudes of the three rings (in a range of around 12 to 25 degrees). The derived latitudes, on the other hand, do not vary much from each other (within a range of only $3^\circ$ to $4^\circ$). The spread in longitudes might be the result of longitudinal electron density variations that result in a varying brightness distribution in the image plane or any physical intrinsic characteristics in the morphology of the flux rope, which is reflected in the results of the manual tracking of the ring fronts. Therefore, under these assumptions, the independent characterization of the three rings indicates that they seem to propagate as a single, coherent structure of the CME.

%---------------------------------------------
\subsection{Flux-rope Morphology}\label{Section_3.2}

The bright and dark nested ring intensity structure observed within the CME suggests a variation of the electron density within the flux rope and possibly the magnetic field morphology. To examine this structure, we characterize the morphology and evolution of the flux rope in 3D space with the help of forward modeling of the outer front of the shell in Section~\ref{sec:GCS} and the observed intensity structure through the cross-sectional morphology of the MFR in Section~\ref{sec:cross-section}.

\subsubsection{Reconstruction of the CME Morphology from multi-point Imaging} \label{sec:GCS}

\begin{figure}[ht]
\begin{center}
\includegraphics[width=15cm]{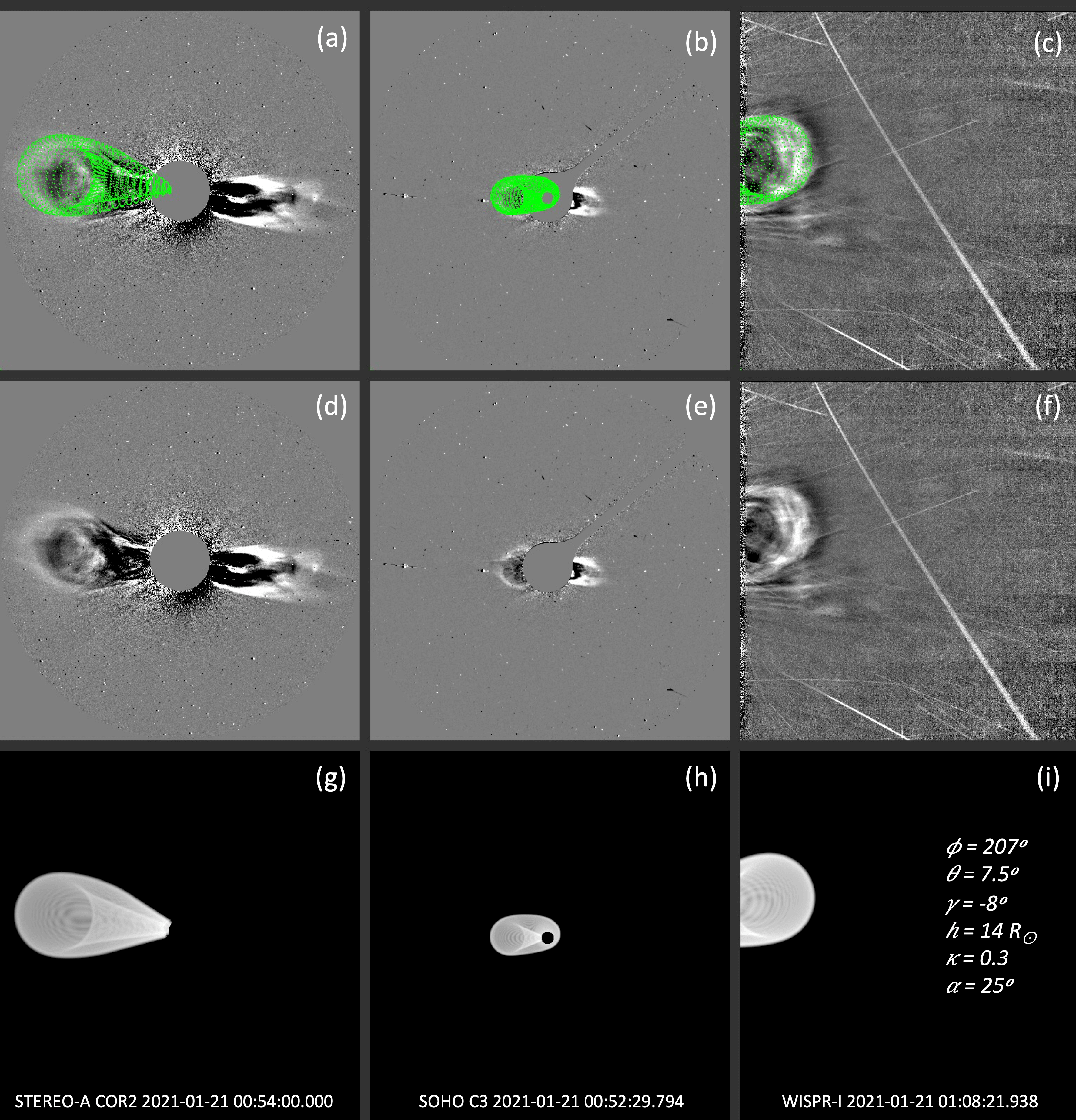}
\end{center}
\caption{GCS reconstruction from background-subtracted ST-A/COR2 (FOV: $2.5$ to $15\ R_\odot$), LASCO/C3 (FOV: $4$ to $30~R_\odot$), and WISPR-I LW (FOV: $9$ to $70~R_\odot$) images chosen close to 01:00 UT on 21 January 2021. Note a second CME in the FOVs of COR2 and C2 from the Western limb (right) of the Sun (not modeled here). The CME on the Eastern limb (left) corresponds to the CME under study and the one observed by WISPR. The flux rope fit of the LASCO C3 coronagraph (panels (b) and (h)) shows its longitudinal extent on the far side of the Sun consistent with the SOHO S/C's viewing geometry and the CME trajectory as discussed in Figure~\ref{Figure_1}(a). The fitting parameters used for the reconstruction are shown in panel (i).}\label{Figure_7}
\end{figure}

Using images from ST-A/COR2, LASCO/C3, and WISPR-I, we conduct forward-modeling of the event from 20 January 2021, 17:00~UT, to 21 January 2021, 06:00~UT to validate the assumption made in Section~\ref{Section_3.1} that the tracked features propagate radially at a constant speed and to determine a set of parameters for interpreting the observed ring structure. We employed the Graduated Cylindrical Shell \citep[GCS;][]{Thernisien_2009, Thernisien_2011} model to characterize the CME's morphology, position, and kinematic evolution. Figure~\ref{Figure_7}(a--c) illustrates the GCS reconstruction from the the three viewpoint observations with the fit of the flux rope shown in green, referred to as the croissant model hereafter. The panels (d--f) display the same snapshots without the croissant models, and panels (g--i) with the best-fit rendered synthetic white-light images of the CME. 

The parameters of this reconstruction include the longitude, $\phi$ = $207^{\circ}$, and latitude, $\theta$ = $7.5^{\circ}$, of the source region of the flux rope in the Stonyhurst coordinate system, the tilt angle of the source region's polarity inversion line, $\gamma$ = $-8^{\circ}$, the height of the leading edge or front of the croissant model, $h$ = $14~R_\odot$, the aspect ratio of the croissant model, $\kappa$ = 0.3, and the half angle between the croissant legs, $\alpha$ = $25^{\circ}$. By conducting this reconstruction several times during the CME propagation in the analyzed time period, the inferred speed of the CME front is $\sim285~km\,s^{-1}$. These parameters agree with those obtained by \cite{Braga_2022}, who noted that after 06:00~UT on 21 January 2021, the gradual deformation of the CME precluded its reliable modeling.

%------------------
%----------------- 

\renewcommand{\thefootnote}{\alph{footnote}}
\setcounter{footnote}{0}
Each of the parameters obtained for the outer ring front, $O_{front}$ from the tracking method of Section~\ref{Section_3.1}, outlined in Table~\ref{table1} and the ones derived from this forward modeling analysis (that mainly fits the outer overall front of the CME) are in good agreement as shown below: longitude ($207^{\circ}$ vs.\ $199.3\pm5^{\circ}$), latitude ($7.5^{\circ}$ vs.\ $7.4\pm2^{\circ}$\footnote{\label{note1}This error is due to the stepping size used in the method, with 2 degrees for latitude and $25~km\,s^{-1}$ for speed, respectively.}), radial distance or height of the croissant front ($10.5~R_\odot$ at 23:00 UT on 20 January 2021 to $18.5~R_\odot$ at 04:00 UT on 21 January 2021 vs.\ $11.5\pm 0.35~R_\odot$ at 23:30 UT on 20 January 2021 to $18.5\pm1.4~R_\odot$ at 04:30 UT on 21 January 2021) and speed ($\sim285~km\,s^{-1}$ vs.\ $\sim270.8\pm 25~km\,s^{-1}$\footref{note1}), respectively. The GCS results, which show no acceleration in the CME propagation during the analyzed time duration, confirm the validity of the assumption of constant speed made in Section~\ref{Section_3.1} and hence validate the tracking methodology.

\subsubsection{Cross-sectional Representation of the MFR:}\label{sec:cross-section}

\begin{figure}[ht]
\begin{center}
\includegraphics[width=15cm]{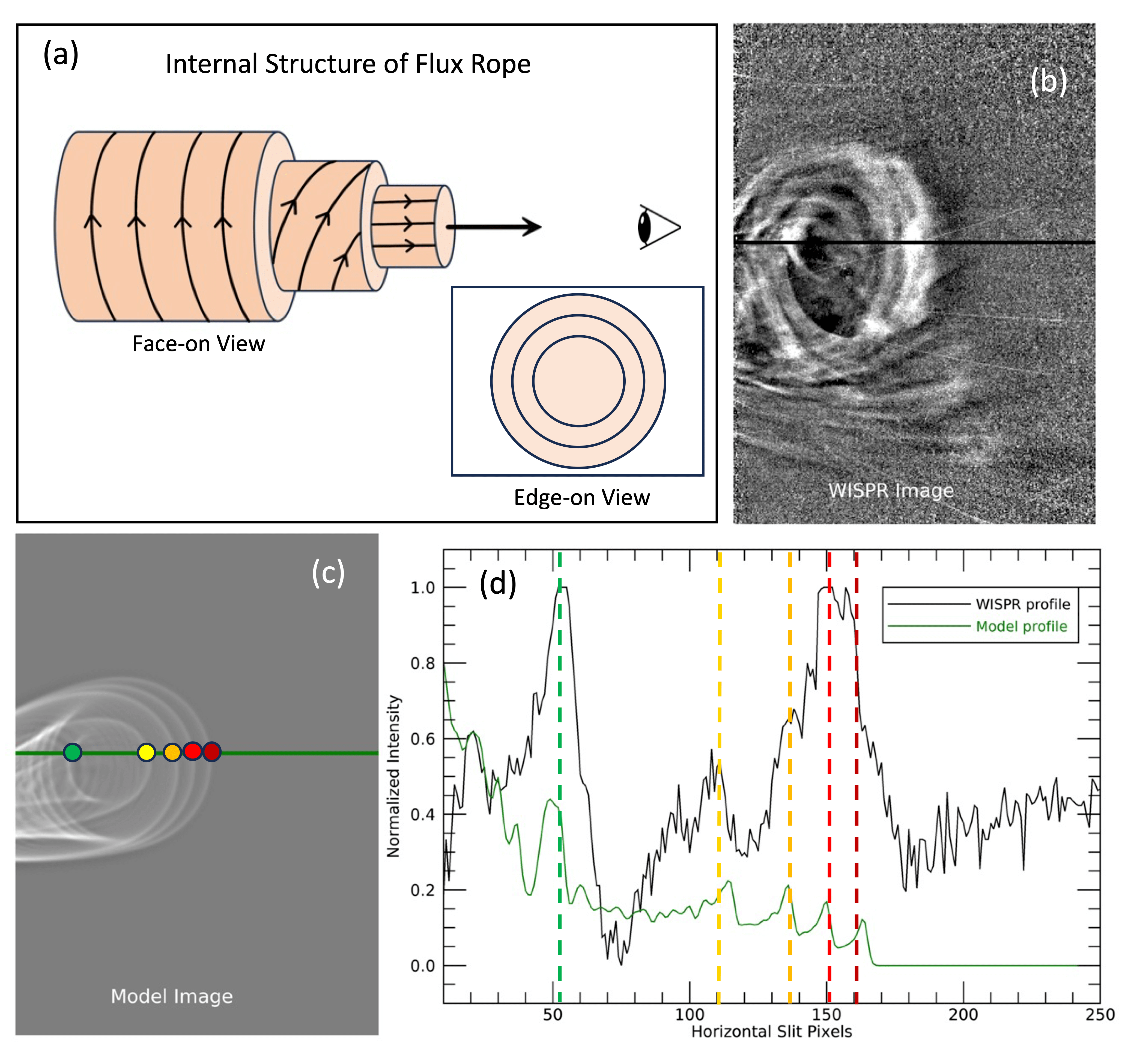}
\end{center}
\caption{Cross-sectional representation of the MFR: (a) A cartoon of the internal structure of a magnetic flux rope showing the orientation of the magnetic field at different (concentric) flux surfaces varying from axial at the core to azimuthal at the outermost surface. The two-dimensional view, viewed from an observer on the right looking along the axis of the flux rope, is shown at the bottom of the panel. (b) Running difference with unsharp masking applied to the LW-processed WISPR image at 04:04 UT on 21 January 2021. This processing was applied to retain the contrast of the finer details of the image. (c) The model image generated from the GCS reconstruction with multiple croissant configurations representing each flux surface mimicking the cartoon model in panel (a). (d) Intensity profiles of the horizontal slits taken on the WISPR and model images, in black and green colors, respectively. Each feature denoted by the colored circles in panel (c) corresponds to the peaks in the intensity profiles (marked by the colored vertical lines).}\label{Figure_8}
\end{figure}

Studies based on magnetic cloud in-situ measurements of interplanetary CMEs have modeled the magnetic morphology of MFRs as consisting of concentric magnetic flux surfaces \cite[][]{Russell_1979, bothmer_1997,bothmer_1998, Hiromitsu_2004} as pictured in Figure~\ref{Figure_8}(a). A MFR is a structure that comprises strands of twisted magnetic field lines wrapping around its central field line or axis. The twist in the flux rope is such that the field orientation changes from azimuthal at the surface to axial at the core. When viewed edge-on (i.e., with the LOS along the axis of the CME flux rope) in white-light observations, this three-dimensional structure can exhibit nested ring-like patterns indicating a ``cross-sectional'' view as shown in the inset box labeled ``edge-on view.'' The closer the field lines are to the axis of the flux rope, the tighter the rings that they form.

In Figure~\ref{Figure_8}(b), we display a snapshot of the flux rope as observed by WISPR-I on 21 January 2021 at 04:04~UT. We notice that the outer rings seem to be non-concentric to each other with an offset in their centers, as pointed out in Section~\ref{Section_3}. To reconstruct a system of flux surfaces, we create its proxy by fitting multiple croissants with the GCS modeling method (multi-croissant configuration). 
 
A planar cut taken at the leading edge of the croissant model perpendicular to the axis of the croissant model results in a circular annulus (similar to the edge-on view of Figure~\ref{Figure_7}(g)), which essentially shows the cross-section of the flux rope at the leading edge. From Equations 28 and 30 of \cite{Thernisien_2011}, and from Figure 1 within that study, the heliocentric distance of the center of the circular annulus ($OC_1$) is given by
\begin{equation}\label{eq3}
    OC_1 = \frac{OH}{1+\kappa},
\end{equation}
where $OH$ is the heliocentric height of the croissant's leading edge, and $\kappa = sin(\delta)$ is the aspect ratio of the croissant model, $\delta$ being the half-angle of a conical leg of the croissant. 

%------------------
%-----------------

For the multi-croissant configuration, we consider four independent croissant models; one for the core ring and three for all the visible outer rings, as in the observations. We first fit the multiple croissants in a concentric configuration by fixing their centers at one location. To achieve the best fit, we needed to check the interplay between the three essential morphological parameters: aspect ratio, height, and center. The heights of the croissant models are set according to the fronts of each ring in the WISPR-I image. From Equation~\ref{eq3} and from the fit on the outermost ring conducted in Section~\ref{sec:GCS}, where $\kappa = 0.3$ and the height of the leading edge $OH = 18.5~R_\sun$ at 04:04~UT, the center of the circular annulus is determined as $OC_1 = 14.23~R_\sun$. Keeping this center constant and utilizing the heights of each of the leading edges of the croissants, we determine the aspect ratios of all the rings needed to generate a concentric configuration. The synthetic white-light image resulting from this concentric configuration does not closely align with the observed image, which suggests that an offset in the shapes and an offset of the centers of the multi-croissant configuration is needed to better match the observation.

As inferred from the WISPR-I snapshot, the shape and co-location of the core ring appear to be different from the three outer rings. In addition, the outer rings seem to be non-concentric to each other, with an offset in their centers, as discussed previously, therefore creating an eccentric circular structure. Hence, to fit this non-concentric configuration, we keep the aspect ratio of the outer rings constant, fine-tune the heights that were obtained previously, and fit the core ring individually to match the observed ring structure. This will eventually lead to an offset in the center of the rings. For the three outer layer rings, at the best fit, we find $\kappa = 0.26$, $OH = [18.0, 17.1, 16.4]~R_\odot$; and for the core ring, $\kappa = 0.14$, $OH = 15.2~R_\odot$. Figure~\ref{Figure_8}(c) shows the reproduced synthetic white-light image obtained from this multiple croissants fitting.

Each GCS croissant shell has density distributed on the surface, where the planar cut perpendicular to the croissant's axis gives a Gaussian profile peaking at the surface and falling off on either side of the surface. To assess the modeling, we examine the intensities along the horizontal slits of three-pixel width taken on both the WISPR (black line) and model croissant (green line) images in Figure~\ref{Figure_8}(b) and (c). The respective averaged over three-pixel intensity profiles are displayed in Figure~\ref{Figure_8}(d) in black and green curves. The vertical dashed lines denote the location of the intensity peaks, which correspond to the features marked with the same color-coded circles in panel (c). We note a direct one-to-one correlation in the locations of the intensity peaks of the modeled features and the observed counterparts. The overall radial fall-off of the intensity of the model cut is due to the superposition of the uniform density of each croissant as we move radially outward in the model image. Note that the image intensity does not show a radial fall-off, which is because the LW processing of the images removes the smooth intensity gradient of the background signal. To summarize, this analysis shows that an edge-on view of the cross-sectional representation of a flux rope schematized as in panel (a) is a plausible representation to explain the appearance of the nested structure in the images. The integrated LOS white-light emission would, therefore, be a signature of the electron density of the plasma trapped within the flux surfaces. 

%---------------------------------------------
\subsection{Visibility of the Nested Ring Structure}\label{Section_3.3}

{As mentioned in Section~\ref{Intro}, a question that arises from the current observations is why most CMEs with flux rope structures do not exhibit discernible nested ring patterns. To examine the plausible causes for the origin of such structure, the magnetic field topology of the CME source region, along with the characterization of its early development in the low corona, is needed. As this CME originated on the far side of the Sun, as seen from Earth and ST-A, the presence of the potentially associated pre-eruptive prominence-cavity system (i.e., a pre-existing flux rope) and the solar source region could not be observed. When the viewing conditions are appropriate, the pre-existing flux rope can be observed as nested patterns in coronal cavities at the solar limb, as discussed in Section~\ref{Intro}. The far-side location of the source region of this event makes it difficult to determine the correspondence between the coronal cavity and the flux rope before and after the CME initiation. Moreover, a statistical inspection of CME events exhibiting nested rings cannot be made either due to insufficiently robust datasets of such cases. Most of the observations prior to Parker launch, except the rare ones through the total solar eclipse \cite[e.g.,][]{Boe_2021}, lack the resolution or vantage point necessary to capture the finer details of CMEs, such as this nested ring structure.

Therefore, we focus on analyzing the presence of the rings mainly in terms of instrumental and observer's perspective effects. If we presume that the nested structures are present in most CMEs, understanding whether the instrument can resolve them is essential. Figure~\ref{Figure_9} shows snapshots of the event as recorded by ST-A/COR2 and WISPR-I. Even though LASCO/C2 and C3 instruments also observed the CME, those observations are not considered here as the nested structure was not clearly visible due to the CME's trajectory towards the back of the Sun as seen from Earth (see Figure~\ref{Figure_1}). In panels (a) and (b), we display the same ST-A/COR2 frame processed in two different ways to check whether the inner structure of the CME and its diffuse front could be revealed. Panel (a) displays a snapshot where the previous image has been subtracted (standard running-difference procedure applied to highlight the development of faint structures). Panel (b), on the other hand, has been processed following the same procedure as the one applied to the LW-processed WISPR-I images. Both snapshots show the circular shape and the irregular cavity region of the CME. Note that the LW-processed snapshot does not exhibit the artifacts that result from a running-difference scheme, i.e., the darker areas behind the bright fronts, which are the result of the corresponding K-corona structures in the image that have been subtracted. None of the processes, however, was able to resolve the complexity of the ring structure and the cavity as seen in WISPR-I images. For easy comparison, a WISPR-I LW-processed snapshot is shown in panel (c). The time of the WISPR-I snapshot was chosen so that the CME extent captured was similar to that of the ST-A/COR2 frame.

%------------------
\begin{figure}[ht]
\begin{center}
\includegraphics[width = 16cm]{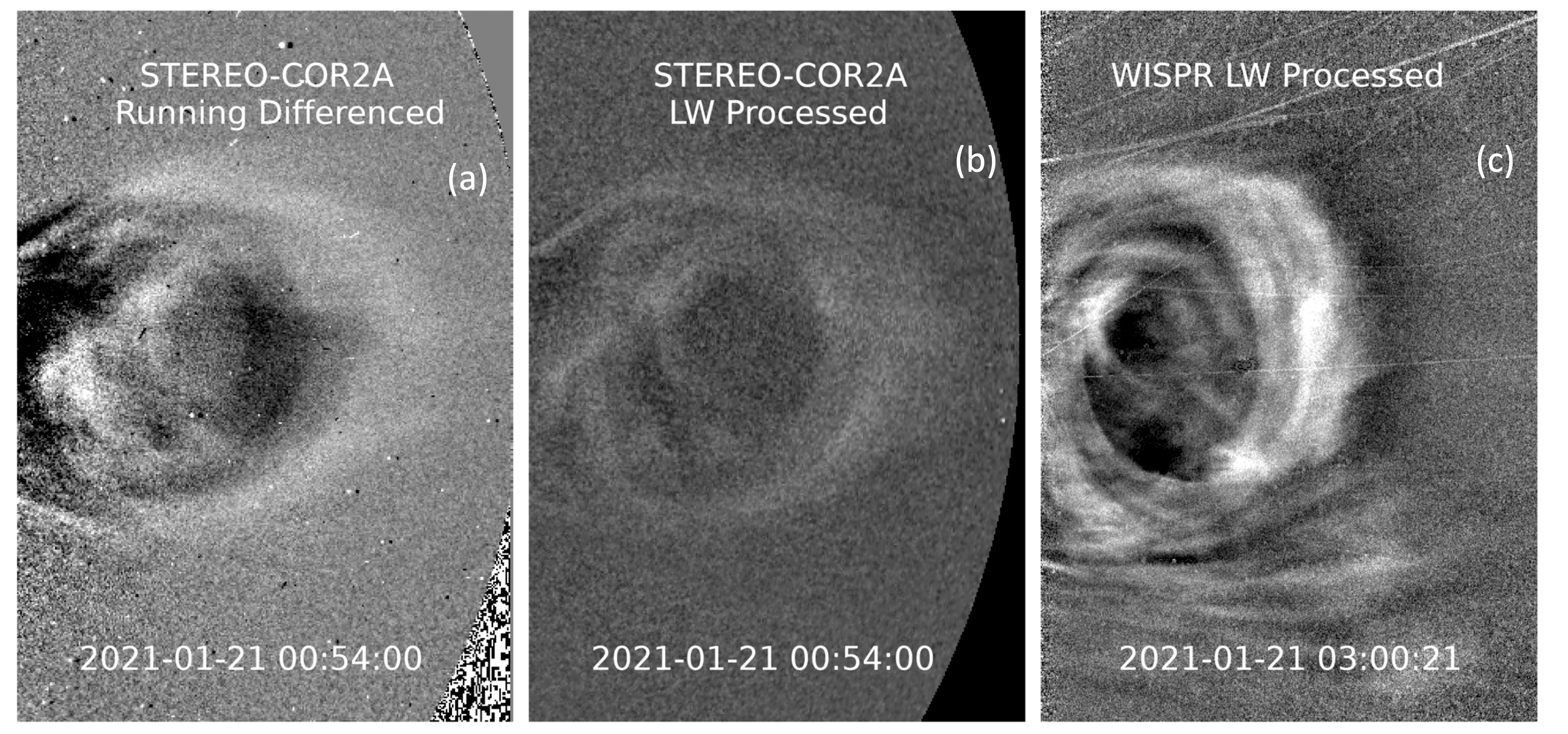}
\end{center}
\caption{Comparison of the CME feature as observed by ST-A/COR2 and WISPR-I. (a) ST-A/COR2 snapshot of the event at  00:54:00 UT on 21 January 2021 is processed following a running difference scheme (i.e., with the preceding frame subtracted). (b) Same ST-A/COR2 snapshot but processed with the LW algorithm. (c) WISPR-I snapshot (processed with the LW algorithm) obtained at a later time, 03:00:21 UT, to match the extent of CME features seen in the ST-A/COR2 image. Both ST-A/COR2 snapshots are mirrored to have the same perspective as the WISPR-I snapshot. }\label{Figure_9}
\end{figure}
%-----------------

Next, we speculate on the factors that might influence the visibility of the ring structure from the two instruments. Both instruments have similar viewing geometry of the CME from opposite sides (see Figure~\ref{Figure_1}). This configuration allows both instruments to observe the CME structure from an edge-on perspective. The spatial resolution of ST-A/COR2 and WISPR-I are about $30~arcsec$ and $22~arcsec$, respectively, in $1~au$ equivalent quantities \citep{Vourlidas2016}, which correspond to $22,000~km$ and $16,000~km$, respectively. Since the speed of the features is ${<}$ $300~km\,s^{-1}$, in the time-lapse of the instrumental exposures ($6.0~s$ and $42.24~s$ for ST-A/COR2 and WISPR-I, respectively), they move less than the distance resolved by each instrument, and hence blurring can be discarded as a factor.

The main contribution to the recorded signal in both instruments is from the F-corona, which results from the integration along the LOS of the photospheric light scattered by the circumsolar dust. ST-A is located at $\sim0.96~au$, while PSP was between $\sim0.17$ and $\sim0.18~au$ during the first five hours of the CME propagation in the WISPR-I FOV. Moreover, the elongations of the LOS that capture the rings are different from the perspective of the two S/C. For these rings, which were observed between $10$ and $20~R_\odot$ (as in Figure~\ref{Figure_6}(c)), the LOS elongations in WISPR-I LOS are above $13.5^\circ$ (up to $108^\circ$ if considering the full evolution of the event across the FOV of the two telescopes), while for ST-A/COR2 they are below $4^\circ$ (starting at $0.7^\circ$). Thus, besides the length of the dust column along the LOS (about $5.4$ times longer for ST-A/COR2 than for WISPR), the smaller elongations play a critical role as the dust scattering efficiency is much larger at small scattering angles, which applies to smaller elongations \citep[see, e.g.,][]{Lamy1986}. Therefore, the relative strength of the K-corona signal compared to that of the F-corona is much smaller in ST-A/COR2 than in  WISPR-I. Hence, after subtracting a proxy of the background signal to reveal the K-corona, the K-features cannot be resolved in as much detail because they are simply buried in the background.

In brief, the prevalent absence of nested rings in white-light coronagraph observations from about $1~au$ might be essentially due to the observer's location and viewing geometry and, certainly, the instrument's design to resolve such structures. In other words, we argue that the spatial resolution and sensitivity of the instrument, along with the length of the dust column through the LOS, are the main factors that play a role in the degree of detail that can be resolved. Therefore, the closer proximity of Parker to the CME under study contributes to an ideal viewing geometry for WISPR and hence becomes the predominant factor for resolving the fine internal structures of this CME.

%---------------------------------------------
\section{Discussion and Conclusions}\label{Section_4}

In this paper, we examine the nested ring CME structure and its evolution observed by the Parker/WISPR imager on 20--21 January 2021.  Other viewpoint observations from ST-A and LASCO coronagraphs are an added advantage to investigate this CME, especially those from ST-A/COR2, for determining the geometrical structure and propagation of the nested ring structure before it undergoes deformation.

This CME exhibits additional density features, such as the inner dip of the flux rope (`U' shaped), the herringbone pattern below the inner dip, and the non-uniform brightness along the front of the CME, as shown in Figure~\ref{Figure_3}. Depending on the projection, these features can be explained by the superposition of the legs and other details of the flux rope. The topology of a CME, as viewed in white-light observations, depends on the orientation of the CME flux rope, where a small writhe to the flux rope may introduce an apparent density substructure in projection \citep[e.g.,][]{CremadesB2004, Vourlidas_2013}. This has also been demonstrated by \cite{THoward_2017}, which used a simple 3D geometric model of a cylindrical tube with bright strands to effectively reproduce a white-light CME as a twisted bundle of magnetic field lines forming a flux rope. As the reader will notice, the animation of their 3D geometrical model \citep[Figure 13 in][]{THoward_2017} is very instructive in interpreting the nested ring structure observed in the WISPR-I images in our study. In particular, at an azimuthal rotation of the model of AZ=$92^\circ$ (a snapshot of which is reproduced here in Figure~\ref{Figure_10}(b), where panel (a) is the initial position of the model), the spiraling strands of the flux rope meet from either side, making a circular pattern that defines the outer shell of the CME, marked by ``1" in panel (b). We believe that each of the locations marked in panel (b), through the line-of-site effects, emulates the additional details observed in the WISPR-I images. Lower down, the superposition of multiple strands at the inner dip of the flux rope (marked by ``2") leads to the `U' shaped brightness, resembling the `U'-shaped pattern observed within the core ring of WISPR-I images. Even farther down, the superposition of both the legs of the CME appears as a bright crisscrossed herringbone pattern (marked by ``3"), also resembling the features in WISPR-I images. In some instances, these bright structures within the CME might be confused with filament material, as discussed in \cite{THoward_2017}. When observed from different viewing angles, this model emphasizes the critical importance of multi-view observations of CMEs in removing ambiguity in understanding their 3D structure.

%------------------
\begin{figure}[ht]
\begin{center}
\includegraphics[width=12cm]{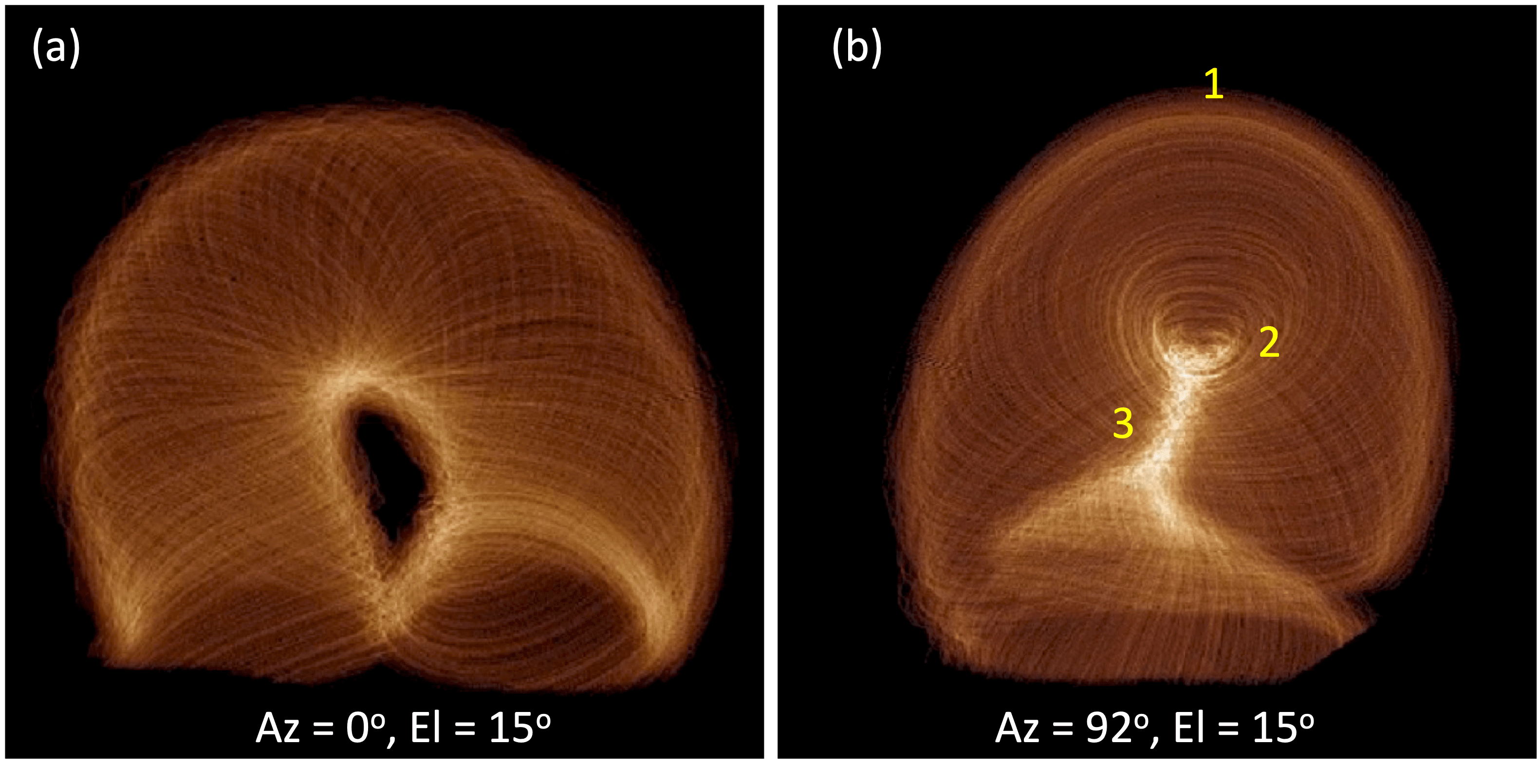}
\end{center}
\caption{Projected 3D model of the simplified twisted and kinked flux rope, as discussed in and adapted from \cite{THoward_2017}: (a) Initial position of the model where the viewing angle of elevation ($El$) is fixed at $15^{\circ}$ and an axis ($Az$) at $0^{\circ}$ (where $El$ is the viewing angle above the $xy$ plane of the three-dimensional Cartesian axes in which the flux rope is located and $Az$ is the azimuthal angle that rotates the axis of the model over $360^{\circ}$). (b) View of the flux rope whose axis is rotated at an azimuthal angle of $92\degree$ relative to the starting position. Refer to the text for details on the features marked by the numbers 1, 2, and 3. Reproduced with permission from \cite{THoward_2017}}\label{Figure_10}
\end{figure}
%-----------------

As shown in \cite{THoward_2017} and numerous other studies, the projection effect plays a critical role, given the optically thin nature of the K-corona signal. The morphology of the flux rope depends on the shear in the source region and its inherent twist, i.e., helicity, of its magnetic field lines. A high degree of shear and helicity in the source region act together to create a highly unstable flux rope, which potentially leads to the eruption as a CME containing a flux rope. When viewed along the axis of the flux rope, the twists in the magnetic field line bundle that forms the flux rope appear as a nested structure in a two-dimensional projection. However, if observed from a direction perpendicular to the axis (i.e., face-on) or from other orientations of the flux rope, the morphological aspect can be different.

Furthermore, Figure~\ref{Figure_11} provides a schematic representation of a projectional view of the flux rope, illustrating how the internal structure of the helical flux rope appears as observed along its axis. When viewed along the axis and through various twists of the flux rope (up to the point where its axis deviates from being parallel to the LOS), the resulting perspective effectively `integrates the projections' of all the twists, similar to looking down the barrel of a twisted straw. The schematic depicts the three-dimensional features of the helical flux rope from an edge-on viewpoint. As shown in the inset box within Figure~\ref{Figure_11}, this perspective reduces the intricate three-dimensional structure of the flux rope to a two-dimensional picture as seen in the images, sometimes making it challenging to fully comprehend in its three-dimensional form.

%------------------
\begin{figure}[ht]
\begin{center}
\includegraphics[width=10cm]{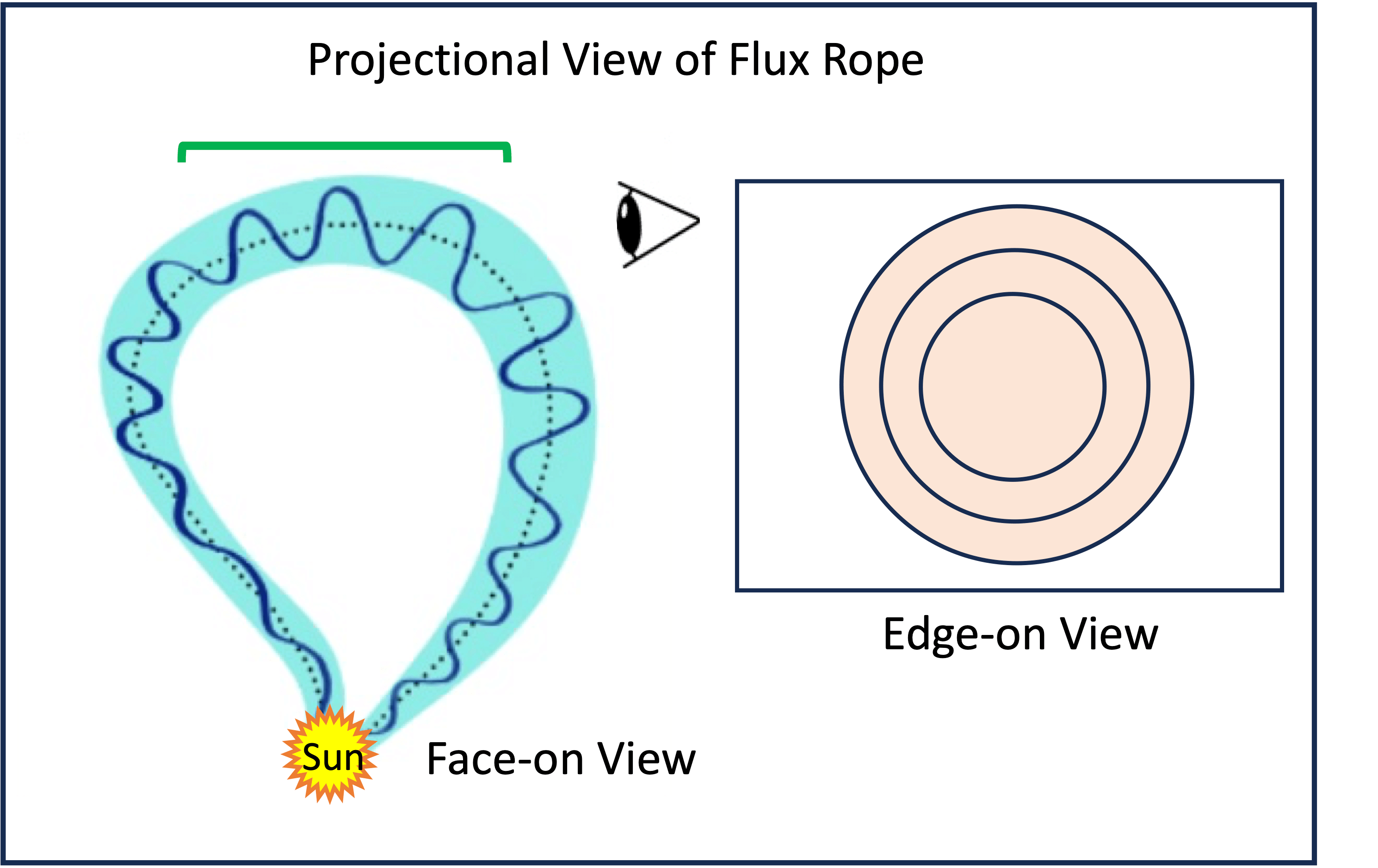}
\end{center}
\caption{Projectional representation of a MFR: The complex, three-dimensional structure of a MFR as it erupts and propagates from the Sun can be visualized through the projections, where each twist within the flux rope in three dimensions can be seen as translated into a two-dimensional representation. The inset box to the right illustrates this projected two-dimensional view.}\label{Figure_11}
\end{figure}
%-----------------

As discussed in \cite{Gibson_2018}, nested ring-like structures in pre-eruption coronal cavities can be explained by reconnective heating at magnetic X-lines or along separatrix layers. One question that remains to be determined is whether the ring-like structures seen in the CME studied here existed in some form before the eruption (pre-existing MFR, as seen in the prominence-nested cavity system), as in the example discussed in \cite{KliemT_2006}, or whether they became regions of enhanced density as a consequence of the eruption \cite[MFR formed on-the-fly;][]{Lynch_2008}, perhaps due to reconnective processes. Unfortunately, this question cannot be answered since observations of the source region before the eruption needed for this were not available in this event because the source region was on the far side of the Sun for the observers with the capabilities to observe the solar disk in EUV wavelengths (e.g., ST-A or missions along the Sun--Earth line). Although Solar Orbiter was also in an ideal location to view this CME's source region with its Extreme Ultraviolet Imager (EUI), observations were not recorded for this date.

The CME nested ring structure can, however, reflect the internal structure of the flux rope, as discussed previously. Each of the nested rings could have been formed in the reconnection process of the erupting magnetic system. When the field lines of a magnetic arcade system sequentially reconnect, the poloidal flux wraps around the central axis (guide field) of a pre-existing flux rope or a flux rope formed on-the-fly. If this wrapping of the flux is intermittent or non-uniform with time and space, probably due to the distinct shear in different locations of the reconnecting arcade of loops or due to distinct reconnection sites within the current sheet, this might lead to the formation of various distinct flux layers of the flux rope. During the reconnection process, the reconnecting field lines, as they converge from different regions, confine plasma with varying densities \citep{Priest_Forbes_2000}. The trapped plasma within the flux surfaces of the newly formed flux rope, in turn, could lead to the nested density structure in the CME after the eruption, as seen in white-light WISPR observations. In addition, the extended period of the eruption, which lasted a few hours, as discussed in Section~\ref{Section_2}, suggests a gradual process of reconnection and eruption at the source region. This may thus lead to clues on the formation of flux ropes characterized by the intermittent wrapping of magnetic flux and density, and this may thus lead to the observed nested structure.

We consolidate the key findings related to the science questions outlined in Section~\ref{Intro} that have been addressed in this study as follows.

\begin{enumerate}[label=\roman*.]
\item Our analysis of the coherence of the ring structure shows that the inner core ring exhibits slight morphological and kinematical differences as compared to the rings in the outer shell of the CME. These differences are, however, within the error range of the measurements. Hence, the overall behavior suggests that the ring structure evolves as one coherent structure as the CME propagates.

\item The observed nested ring structure is attributed to the morphology of the flux rope and its internal substructure, as depicted in the cross-sectional representation of the flux rope. We also discuss various scenarios, including the projectional representation of a MFR, the superposition of the internal structures of a MFR, and magnetic reconnection for the formation and visibility of the nested ring structure. However, understanding the formation of these structures via magnetic reconnection requires modeling the CME current sheet and flux rope formation, which is beyond the scope of this current paper. 

\item We emphasize that appropriate viewing conditions, such as the spatial resolution of the instrument, close observing distance, and optimal viewing angles, are needed to resolve the finer substructures within CMEs. Conducting a statistical analysis to determine the prevalence of nested ring structure in all CMEs and the underlying reasoning behind it is essential but is not possible in this study due to the lack of a robust dataset of observations fulfilling the conditions mentioned above.
\end{enumerate}

In conclusion, this study highlights the unique perspective offered by the Parker mission to observe solar phenomena in white light emission. As shown, the vantage location of Parker helps WISPR reveal the finer and more detailed structure of CMEs, which otherwise appear mostly hidden in observations from 1 au with the existing instruments, and hence discern the electron density features embedded in the magnetic structure of CME flux ropes. These observations show the existence and the evolution of the apparent structural integrity of the flux ropes and their manifestations at large radial distances from the Sun observed by Parker.

%---------------------------------------------------------------------------
\section*{Acknowledgments}

This work was supported by the NASA Parker Solar Probe Program Office for the WISPR program (Contract NNG11EK11I to the U.S.\ Naval Research Laboratory; NRL).
S.B.S. acknowledges the support from George Mason University (GMU) via the NRL contract (N00173-23-2-C603).
M.G.L.,\ P.H.,\ and R.C.C.\ acknowledge support from the Office of Naval Research.
G.S. acknowledges support from the WISPR Phase-E program at APL.
C.R.B.\ acknowledges the support from the NASA STEREO/SECCHI (NNG17PP27I) program and NASA's HGI grants 80NSSC23K0412 and 80NSSC24K1252.
E.P.\ acknowledges support from NASA's PSP-GI (80NSSC22K0349) and HTMS (80NSSC20K1274) programs as well as the PSP/WISPR contract to NRL (under subcontract no.\ N00173-19-C-2003 to PSI).
The National Center for Atmospheric Research (NCAR) is a major facility sponsored by the National Science Foundation under Cooperative Agreement No.\ 1852977.
S.B.S.\ thanks R. A. Howard for his comments, which greatly improved this manuscript. S.B.S.\ also thanks A. Vourlidas for having numerous productive discussions on CME science and for processing the high-cadence ST-A/COR2 data. 
The Parker Solar Probe was designed and built and is now operated by the Johns Hopkins Applied Physics Laboratory as part of NASA’s Living with a Star (LWS) program (contract NNN06AA01C). Support from the LWS management and technical team has played a critical role in the success of the mission.
The Wide-Field Imager for Parker Solar Probe (WISPR) instrument was designed, built, and is now operated by NRL in collaboration with the Johns Hopkins University Applied Physics Laboratory, California Institute of Technology Jet Propulsion Laboratory, University of G{\"o}ttingen (Germany), Centre Spatiale de Li{\`e}ge (Belgium), and University of Toulouse Research Institute in Astrophysics and Planetology (France). The data used in this study are publicly available via the WISPR Data Access on the Wide-Field Imager for Parker Solar Probe Project Website, which can be found at \href{https://wispr.nrl.navy.mil/}{https://wispr.nrl.navy.mil/}.
We acknowledge the SOHO/LASCO and STEREO/SECCHI teams for making the data used in this study available. 

\bibliographystyle{aasjournal} 
\bibliography{NestedCME}

\end{document}